\documentclass[preprint2]{aastex}

\def\arcsec{\hbox{$^{\prime\prime}$}}

\slugcomment{Accepted to AJ}

\shorttitle{HD 15137}
\shortauthors{McSwain et al.}

\begin{document}

\title{Multiwavelength Observations of the Runaway Binary HD 15137
\footnote{Based partly on observations collected at the Observatoire de Haute-Provence (France) and on observations obtained with XMM-Newton, an ESA science mission with instruments and contributions directly funded by ESA Member States and NASA. } }

\author{M. Virginia McSwain\altaffilmark{2}}
\affil{Department of Physics, Lehigh University, Bethlehem, PA 18015}
\email{mcswain@lehigh.edu}

\author{Micha\"el De Becker}
\affil{Institut d'Astrophysique et G\'eophysique, Universit\'e de Li\`ege, FNRS, Belgium}
\email{debecker@astro.ulg.ac.be}

\author{Mallory S.\ E.\ Roberts}
\affil{Eureka Scientific, Inc., Oakland, CA 94602-3017}
\email{malloryr@gmail.com}

\author{Tabetha S.\ Boyajian\altaffilmark{2,3}, Douglas R.\ Gies}
\affil{Department of Physics and Astronomy, Georgia State University,
Atlanta, GA 30302-4106}
\email{tabetha@chara.gsu.edu, gies@chara.gsu.edu}

\author{Erika D.\ Grundstrom\altaffilmark{2}}
\affil{Physics and Astronomy Department, Vanderbilt University, Nashville, TN 37235}
\email{erika.grundstrom@vanderbilt.edu}
             
\author{Christina Aragona\altaffilmark{2}, Amber N.\ Marsh\altaffilmark{2}, Rachael M.\ Roettenbacher\altaffilmark{2}}
\affil{Department of Physics, Lehigh University, Bethlehem, PA 18015}
\email{cha206@lehigh.edu, anm506@lehigh.edu, rmr207@lehigh.edu}

\altaffiltext{2}{Visiting Astronomer, Kitt Peak National Observatory. KPNO is operated by AURA, Inc.\ under contract to the National Science Foundation.}
\altaffiltext{3}{Hubble Fellow}


\begin{abstract}

HD 15137 is an intriguing runaway O-type binary system that offers a rare opportunity to explore the mechanism by which it was ejected from the open cluster of its birth.  Here we present recent blue optical spectra of HD 15137 and derive a new orbital solution for the spectroscopic binary and physical parameters of the O star primary.  We also present the first \textit{XMM-Newton} observations of the system.  Fits of the EPIC spectra indicate soft, thermal X-ray emission consistent with an isolated O star.  Upper limits on the undetected hard X-ray emission place limits on the emission from a proposed compact companion in the system, and we rule out a quiescent neutron star in the propellor regime or a weakly accreting neutron star.  An unevolved secondary companion is also not detected in our optical spectra of the binary, and it is difficult to conclude that a gravitational interaction could have ejected this runaway binary with a low mass optical star.  HD 15137 may contain an elusive neutron star in the ejector regime or a quiescent black hole with conditions unfavorable for accretion at the time of our observations.   

\end{abstract}



\section{Introduction}

While most O- and B-type stars are believed to form in open clusters and stellar associations, some are observed at high galactic latitudes and with large peculiar space velocities \citep{gies1986}.  
These stars were likely ejected from the clusters of their birth, and there are two accepted mechanisms to explain the origin of their runaway velocities.  Close multi-body interactions in a dense cluster environment may cause one or more stars to be scattered out of the region \citep{poveda1967}, and about 10\% are expected to be ejected as binary pairs \citep{leonard1990}.  An alternative mechanism involves a supernova explosion within a close binary, ejecting the secondary due to the conservation of momentum \citep{zwicky1957}.  The resulting neutron star (NS) or black hole (BH) may remain bound to the secondary if not enough mass is lost during the explosion or if a retrograde kick velocity occurs, and a binary fraction of $20-40$\% is predicted for these runaways \citep{portegieszwart2000}.  The observed fraction of binaries among runaways seems consistent with either scenario ($5-26$\%; \citealt{mason1998}).  While both processes do occur in nature, it is not clear which process is dominant.  

Distinguishing between the dynamical ejection and binary supernova scenarios for an isolated runaway star can be nearly impossible.  However, binary runaway systems offer a significant advantage to cluster ejection studies since multiwavelength observations can pinpoint NS companions produced in supernovae.  Optical spectroscopic studies alone offer circumstantial evidence for compact companions, primarily by high eccentricity and low mass function determined from the orbit, but also due to nitrogen and/or helium enrichment and fast rotational velocity that suggest prior mass transfer \citep{hoogerwerf2001}.  A few runaway massive stars are predicted to be associated with radio pulsars \citep{portegieszwart2000}, but so far none have been found \citep{philp1996, sayer1996}.  X-ray emission is a better diagnostic since stellar wind interactions with a NS produce a measurable X-ray excess above the normal, intrinsic emission from shocks in the stellar winds \citep{sana2006, popov2000, lamers1976}.  

HD 15137 is a runaway binary that offers the rare opportunity to diagnose its cluster ejection mechanism.  Its spectral type is O9.5 III(n).  \citet{boyajian2005} found that this runaway single-line spectroscopic binary (SB1) system was likely ejected from the open cluster NGC 654, but the travel time since its ejection, 10 Myr, presents an unusual paradox since it is longer than the expected lifetime of an O-type star.  \citet{boyajian2005} and \citet{mcswain2007a} presented preliminary orbital solutions for the SB1, finding a highly eccentric orbit with a period of about 30 d.  The O-type star has an effective temperature $T_{\rm eff} = 29700$ K and surface gravity $\log g = 3.50$, is located at a distance of 2.420 kpc, and has a fast peculiar space velocity $V_{\rm pec} = 63$ km~s$^{-1}$ \citep{mcswain2007a}.  Such a value of $V_{\rm pec}$ places this SB1 system among the class of runaway O stars.  

The very low mass function suggests a low mass companion in HD 15137, yet so far \citet{boyajian2005} and \citet{mcswain2007b} have tried unsuccessfully to identify the nature of the companion star.  Using doppler tomographic separation with their preliminary orbital solution, \citet{boyajian2005} were unable to identify an unevolved secondary companion in their red optical spectra.  
Furthermore, we simulated 4-body interactions of HD 15137 with the STARLAB package and found a very low probability that dynamical interactions could have ejected a low mass, unevolved star with the O-type primary \citep{mcswain2007b}.  Its relatively short period, eccentric orbit led \citet{boyajian2005} to suggest that the orbit was widened during a supernova event and that tidal circularization has not yet occurred.  The system is not a known X-ray source, so it has been proposed to be a high mass X-ray binary (HMXB) candidate.  Current upper limits of the X-ray flux rule out a stellar wind-accreting NS in the system, and we performed a radio pulsar search with null results \citep{mcswain2007b}.  

In this work, we present new blue optical spectra of HD 15137 in \S 2.  Based on new radial velocities of the system, we argue that the orbital period is somewhat longer than we have previously determined, and we present an improved orbital solution for this SB1.  In \S 3, we present the stellar physical parameters determined by comparing our optical spectra to model spectra.   We describe the first X-ray detection of HD 15137 with the \textit{XMM-Newton} observatory in \S 4.  From fits of the soft X-ray spectra, we place upper limits on the hard X-ray flux from a potential NS companion.  We discuss these results further in \S 5, and summarize the key results in \S 6.


\section{Optical Observations and Radial Velocity Measurements}

We obtained 44 spectra of HD 15137 at the Observatoire de Haute-Provence (OHP) during several observing runs from 2005 October to 2007 November, including 23 spectra obtained over 30 consecutive nights in fall 2007.  We used the 1.52m telescope with the Aur\'elie spectrograph with grating \#3 and the $2048\times1024$ CCD EEV 42-20\#3 detector with a pixel size of 13.5 $\mu$m.  These spectra cover a wavelength range between 4460--4890 \AA~with a resolving power of $R = \lambda/\Delta \lambda \approx 8000$.  The exposure times were typically 30 minutes but sometimes longer depending on weather conditions.  On 2006 October 27, we performed an intense stare at HD 15137 with 12 consecutive exposures, although on most other nights only a single spectrum was obtained.  The Aur\'elie CCD data consist of five images of the object's spectrum obtained using a Bowen-Walvaren image slicer (described in \citealt{gillet1994}). The OHP data were bias corrected, flat fielded, cleaned for cosmic rays, and extracted using the MIDAS software developed at the European Southern Observatory.  We performed wavelength calibration of each spectrum using a ThAr comparison spectrum obtained shortly before or after each observation.  The spectra were rectified using line-free continuum regions and interpolated onto a log wavelength scale using a common heliocentric wavelength grid.  

We also obtained 47 spectra of HD 15137 at the KPNO coud\'e feed (CF) telescope over 35 consecutive nights during 2008 October and November.  We used grating B in third order with the 4-96 order-sorting filter and the F3KB detector.  This instrumental configuration resulted in a wavelength range of 4130--4570 \AA~with $R \approx 9000$ across the chip.  We generally obtained one spectrum of 15 minutes duration each night, except for one night (2008 November 1) when we performed another long stare at HD 15137, collecting 16 spectra over four hours.  We also obtained ThAr comparison spectra about every 2--3 hours throughout each night.  The spectra were bias corrected, flat fielded, and wavelength calibrated using standard procedures in IRAF\footnote{IRAF is distributed by the National Optical Astronomy Observatory, which is operated by the Association of Universities for Research in Astronomy, Inc., under cooperative agreement with the National Sciences Foundation}.  Finally, the coud\'e feed spectra were rectified to a unit continuum using line-free regions and interpolated onto a log wavelength scale using a common heliocentric wavelength grid.

\begin{figure}
\includegraphics[angle=180,scale=0.35]{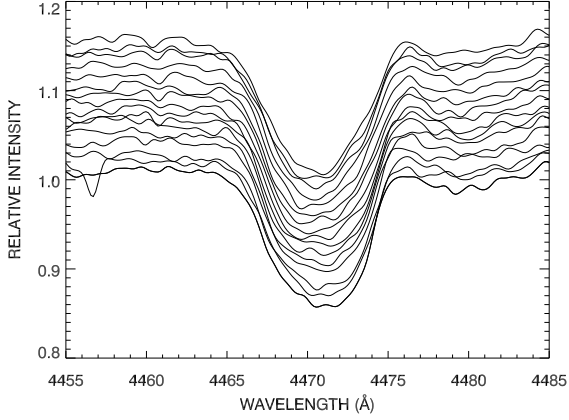}
\caption{
\label{lpv}
Variations in the \ion{He}{1} $\lambda4471$ line profile of HD 15137 during the CF long stare.  The spectra have been smoothed with a Gaussian of FWHM = 5 pixels and are offset vertically for clarity.
}
\end{figure}

The recent optical spectra of HD 15137 show evidence of low amplitude, high frequency line profile variations, especially during the long stares that were performed in each of the runs described above.  The variations in the \ion{He}{1} $\lambda4471$ line are shown in Figure \ref{lpv}.  While the bumps in the line profile are consistent with the noise level of the continuum, the line profile is also non-symmetric and shifting slightly over the long stare.  These short-term variations, likely due to nonradial pulsations, were suggested originally by \citet{boyajian2005} and have been described in more detail by \citet{debecker2008}.  Superimposed on these low amplitude line profile variations, we observed larger radial velocity, $V_r$, variations of each absorption line.  Thus HD 15137 appears to be a single-line spectroscopic binary (SB1) that also exhibits line profile variations.  

To measure $V_r$ of HD 15137, we used a cross-correlation procedure similar to that described in \citet{mcswain2007a}.  However, we used a mean spectrum created from each data set, rather than any single observation, as a reference spectrum for the cross correlation process.  This choice allowed us to minimize the effects of small-scale line profile variations and identify the bulk $V_r$ shifts in the data instead.  Using this mean spectrum, we fit the core of each absorption line with a parabola to determine its absolute radial velocity.  The remaining spectra were then cross correlated with this template to determine $V_r$ for each line.  We used all available strong, unblended absorption lines to measure $V_r$, and rest wavelengths for each line were taken from the NIST Atomic Spectra Database\footnote{The NIST Atomic Spectra Database is available online at http://physics.nist.gov/PhysRefData/ASD/index.html.}.  In the CF data set, the lines used to measure $V_r$ were H$\gamma$, \ion{He}{1} $\lambda\lambda4144, 4388, 4471$, and \ion{He}{2} $\lambda4200$.  In the OHP data set, we used the H$\beta$, \ion{He}{1} $\lambda\lambda4471, 4713$, and \ion{He}{2} $\lambda\lambda4542, 4686$ absorption lines.  

The resulting mean $V_r$ and standard deviation, $\sigma$, are listed in Table \ref{vr}.  In many cases, the measured $\sigma$ probably underestimate the true error in $V_r$ due to the line profile variations.  During both of our long stares at HD 15137, we observed changes in $V_r$ of about 10 km~s$^{-1}$ over only a few hours.  We recommend adding this error of $\pm 5$ km~s$^{-1}$ in quadrature with $\sigma$ for a more representative error in $V_r$.

In our previous spectroscopic studies of HD 15137 \citep{boyajian2005, mcswain2007a}, we proposed an orbital period $P \sim 30$ d for the system.  However, our new $V_r$ measurements exclude such a value of $P$, especially using our data from two observing runs that each took place over $\sim 30$ consecutive nights.  We performed a new period search on all available $V_r$ measurements of HD 15137 using a version of the discrete Fourier transform and CLEAN deconvolution algorithm of \citet*{roberts1987} (written in IDL\footnote{IDL is a registered trademark of Research Systems, Inc.} by A.\ W.\ Fullerton).  In addition, we used the phase dispersion minimization (PDM) algorithm \citep{stellingwerf1978} to compare with the CLEANed power spectrum.  The PDM method is better suited for eccentric orbits that are strongly non-sinusoidal.  

There was no one clear signal that stands out from any of the resulting periodograms, so we inspected each candidate frequency carefully.  We used each proposed period as input into the non-linear, least-squares fitting program of \citet{morbey1974} to solve for the resulting orbital elements.  After ruling out all resulting $V_r$ curves with poor fits and extremely large scatter (and also any $P < 35$ d), we settled upon two acceptable solutions with nearly equally good fits and frequencies 0.018 d$^{-1}$ and 0.015 d$^{-1}$, corresponding to $P = 55.4$ d and $P = 65.2$ d, respectively.

Since our data are often spaced at intervals on the order of 30 d, this produces a sampling frequency $f_s = 0.033$ d$^{-1}$, which is also a signal found by CLEAN (and also corresponds to our earlier proposed period).  Our sampling rate induces an alias frequency $f_a = f_s - f$, where $f$ is the true signal frequency.  Since no significant aliasing would occur if the Nyquist condition is satisfied, $f < f_s/2$, we attribute the lower frequency as the alias and adopt the true orbital frequency $f = 0.018$ d$^{-1}$.  Thus we adopt an orbital period $P = 55.4$ d for HD 15137.  We note that a long period aliasing effect also appears prominently in our results; the strong peaks near 0.0015 d$^{-1}$ and 0.007 d$^{-1}$ found by CLEAN are the result.  

As further verification of our adopted orbital period, we tried manipulating our collection of $V_r$ data in several ways to test whether the period search results remained consistent.  We performed a period search with the CF dataset omitted, finding two potential periods of 57 d and 67 d.  When we instead omitted the OHP data from the collection, a period of 52 d stands alone as the most prominent signal.  Finally, we repeated the period search an additional ten times, randomly removing 20\% of all of the $V_r$ points in each trial.  Prominent signals remain consistently at 55 d and 65 d.  These results provide further support that the timescales of 55 d and 65 d are good candidates for the orbital period, with the former preferred.  However, we recommend that both values be considered with caution since low frequency trends also appear to contribute to the temporal behavior of the radial velocities.  The corresponding orbital parameters should be treated with caution as well.  

\begin{figure}
\includegraphics[angle=180,scale=0.35]{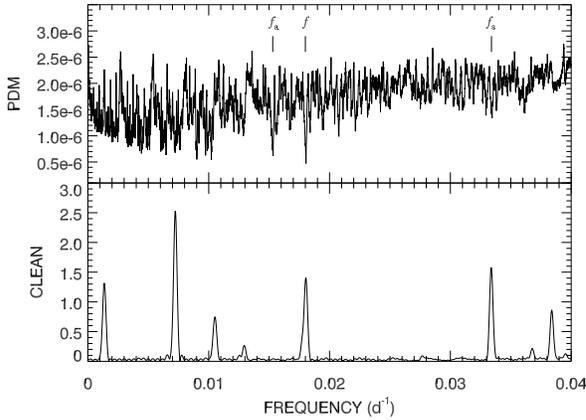}
\caption{
\label{periodogram}
Frequency search results using PDM (top) and CLEAN (bottom) methods.  The sampling frequency $f_s$, alias frequency $f_a$, and true signal frequency $f$ are also marked in the top panel.  
}
\end{figure}

Finally, we inspected the complete $V_r$ dataset for systematic velocity differences by repeating the orbital fit for each set separately, allowing only $V_0$ to vary.  The $V_r$ offset was determined by bringing each $V_0$ into agreement with $V_0$ from the CF set.  Thus we offset the OHP measurements by $-4.97$ km~s$^{-1}$, measurements from \citet{boyajian2005} by $+2.7$, and measurements from \citet{mcswain2007a} by $+2.4$.  Note that the \textit{unaltered} OHP measurements are presented in Table \ref{vr}.  We then repeated the period search and orbital fit with the corrected $V_r$, which improved our orbital solution significantly.  
Figure \ref{periodogram} shows the results from the final period searches.  Note that for the CLEAN method, peaks in the signal indicate likely frequencies, while for the PDM method the minima indicate more probable frequencies.  Our systematic $V_r$ correction removed $f_a$ in the CLEAN search, but it remains present in the PDM periodogram. 

\begin{figure}
\includegraphics[angle=180,scale=0.35]{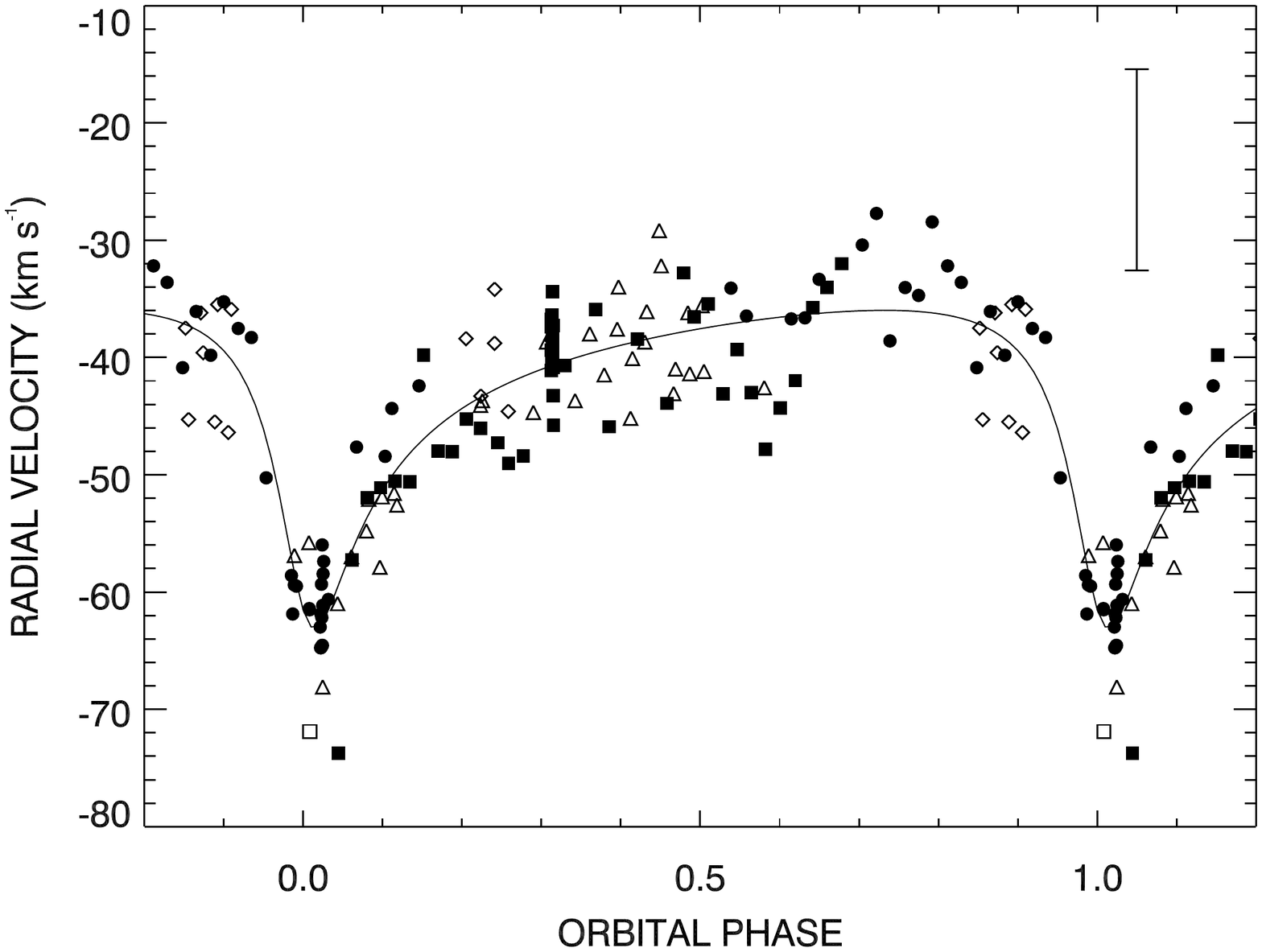}
\caption{
\label{vrcurve}
Radial velocity curve of HD 15137.  The single point from \citet{conti1977} is plotted as an open square, points from \citet{boyajian2005} as open triangles, points from \citet{mcswain2007a} as open diamonds, points from the OHP in this work as filled circles, and points from the CF in this work as filled squares.  Our long stares were performed at $\phi \approx 0.0$ and $\phi \approx 0.3$.  A typical error bar, assuming $\sigma = 7$ km~s$^{-1}$ in addition to the intrinsic $V_r$ error of $\pm 5$ km~s$^{-1}$ due to rapid line profile variations, is also shown.  
}
\end{figure}

\begin{figure}
\includegraphics[angle=180,scale=0.35]{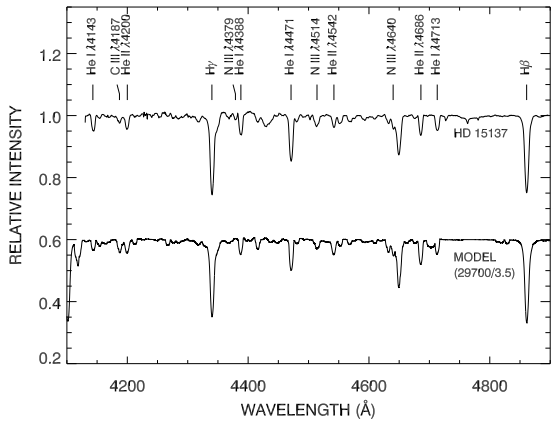}
\caption{
\label{meanspec}
Mean spectrum of HD 15137 from our CF observations ($\lambda < 4450$ \AA) and from OHP ($\lambda > 4450$ \AA).  The CF spectrum has been smoothed for presentation.  A model spectrum interpolated from the Tlusty OSTAR2002 grid \citep{lanz2003}, with $V \sin i = 258$ km~s$^{-1}$, $T_{\rm eff} = 29700$ K, and $\log g = 3.5$, is shown for comparison.  
}
\end{figure}

We present the final orbital solution in Table \ref{orbit}, and the corresponding radial velocity curve in Figure \ref{vrcurve}.  The mean spectrum of HD 15137 from our CF and OHP runs, shifted to its rest wavelength according to our orbital solution, is shown in Figure \ref{meanspec}.  
The final orbital solution of HD 15137 indicates a highly eccentric ($e = 0.62$) binary with a low velocity semiamplitude ($K = 13.6$ km~s$^{-1}$).  These orbital elements are very similar to our preliminary orbit \citep{mcswain2007a}, although the period is quite different.  While the O9.5 III primary is likely a massive star, the very low mass function of this binary suggests a low mass companion.  Based on the runaway nature of the binary, the eccentric orbit, and the low mass function, \citet{boyajian2005} proposed a supernova ejection scenario and a NS companion in HD 15137, even though the system was not a known X-ray source.  They proposed that HD 15137 may be a ``quiet'' HMXB, too widely separated for the NS to accrete a significant mass of stellar winds to produce the bright X-ray flux commonly associated with X-ray binaries \citep{liu2006}.  To investigate this quiet HMXB scenario, we describe \textit{XMM-Newton} observations of the system below in \S 4.  


\section{Physical Parameters from Spectral Models}

In \citet{mcswain2007a}, we measured the stellar parameters of HD 15137 using the Tlusty OSTAR 2002 grid of line-blanketed, non-LTE, plane-parallel, hydrostatic atmosphere model spectra for O-type stars \citep{lanz2003}.  The OSTAR2002 grid uses a microturbulent velocity $V_t = 10$ km~s$^{-1}$, and we assumed solar abundances for the star.  In our earlier paper we found $V \sin i = 234$ km~s$^{-1}$, $T_{\rm eff} = 29700$ K, and $\log g = 3.5$ for HD 15137 using fits of the \ion{He}{2} line profiles.  
Given that we have obtained higher resolution blue spectra for a new comparison, we repeated the spectral modeling using the Tlusty OSTAR2002 grid \citep{lanz2007}.  

We measured $V \sin i$ by comparing the observed \ion{He}{1} $\lambda\lambda 4143, 4388, 4471, 4713$ line profiles in the mean CF and OHP spectra to the OSTAR2002 model line profiles convolved with a limb-darkened, rotational broadening function and a Gaussian instrumental broadening function.  We determined the best fit over a grid of $V \sin i$ values by minimizing the mean square of the deviations, rms$^2$.   The formal error, $\Delta V \sin i$, is the offset from the best-fit value that increases the rms$^2$ by $2.7 \, \rm rms^2$/$N$, where $N$ is the number of wavelength points within the fit region.  We measured a new value of $V \sin i = 258 \pm 20$ km~s$^{-1}$, consistent with our earlier result.  

We measured $T_{\rm eff}$ by comparing the observed \ion{He}{2}:\ion{He}{1} equivalent width ratios to the equivalent width ratios in the broadened model spectra \citep{walborn1990}.  The OSTAR2002 grid overlaps with the Tlusty BSTAR2006 grid over the range of $T_{\rm eff}$ and $\log g$ appropriate for a late O-type star, although the BSTAR2006 grid uses a lower $V_t$ (2 km~s$^{-1}$ for our values of $\log g$; \citealt{lanz2007}).  To consider whether lower values of $V_t$ may be appropriate for HD 15137, we also compared our spectra to both grids over the appropriate range in $T_{\rm eff}$ and $\log g$.  We measured equivalent widths, $W_\lambda$, using numerical integrations over each line profile, and we estimate an error of 10\% in each observed $W_\lambda$ due to errors in the continuum placement and intrinsic noise.  By fixing $\log g$ at each of three possible values ($\log g = 3.25$, 3.5, and 3.75), we determined a nominal value of $T_{\rm eff}$ from our observed $W_\lambda$ ratios by interpolation with the model $W_\lambda$ ratios.  The resulting values of $T_{\rm eff}$ and $\log g$ are listed in Table \ref{specmodel}.  Using the mean $T_{\rm eff}$ for each value of $\log g$, we then compared the broadened model spectrum to our mean observed spectra over the full wavelength range and determined a reduced $\chi^2$ from the difference.  The resulting values of reduced $\chi^2$ for each fit are listed in Table \ref{specmodel}.  Based upon our renewed inspection of the \ion{He}{1} and \ion{He}{2} line strengths, our new measurements for $T_{\rm eff} = 29700$ K and $\log g = 3.5$ agree perfectly with our earlier study.  We estimate the error in $T_{\rm eff}$ to be 1700 K based on the standard deviation of the three measured values, and we estimate an error in $\log g$ to be 0.25 dex due to the spacing of the grids of model spectra.  

The final broadened model spectrum from the OSTAR2002 grid is presented in Figure \ref{meanspec}.  Compared to the model spectrum, HD 15137 has a slightly weak \ion{C}{3} $\lambda4187$ line and strong \ion{N}{3} $\lambda\lambda4379, 4514$ lines.  This may be evidence that the atmosphere is enriched with CNO-processed gas.  Finally, we note that there is no evidence for an optical companion present in our observations based upon the excellent agreement of our observations and this model of an isolated O-type star.


\section{XMM-Newton Observations}

We observed HD 15137 with the \textit{XMM-Newton} observatory on 2008 August 3, observation ID 0553810201, for approximately 20 ks.  Based on our proposed orbital solution, these observations took place at orbital phase $\phi = 0.69$.  The field of view was centered on HD 15137 for a direct on-axis view.  The three EPIC cameras were operated in Full Frame mode with the medium optical filter, and the Optical Monitor was turned off during the observation due to the brightness of the star.  The RGS instruments were operated in Standard Spectroscopy mode, but due to the faintness of the source the RGS data were not used.

The EPIC Observation Data Files (ODFs) and event lists were provided by the standard XMM Pipeline Processing System.  Using the XMM-Science Analysis System (SAS) version 7.1.2, we filtered the event lists using the standard cutoff of 0.35 counts~s$^{-1}$ for the MOS cameras and 5 counts~s$^{-1}$ for the pn camera using the \textit{evselect} command to exclude times of high particle background.  The resulting effective exposure times (good time intervals) for each camera are listed in Table \ref{xmmjournal}.  

We extracted the source spectra from the event lists using a circular region with radius of 25\arcsec~for the MOS cameras and 32\arcsec~for the pn camera.  To extract the background spectrum for each camera, we used a partial annulus region with inner and outer radii of 80\arcsec~and 110\arcsec, respectively, centered on the source.  Because another possible weak X-ray source was detected in this annular region, we excluded the portion of the annulus near that source.  The resulting count rates for the source and background regions are listed in Table \ref{xmmjournal}.  As a weak X-ray source, we found no indication of pile-up in the observation.  

\begin{figure}
\includegraphics[angle=0,scale=0.35]{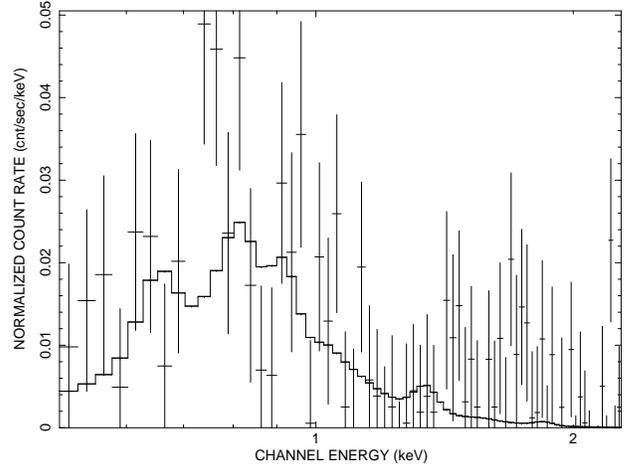}
\caption{
\label{xmmspec}
\textit{XMM-Newton} pn spectrum of HD 15137.  The spectrum has been fit in XSPEC using a warmly absorbed thermal model, \textit{wabs(apec)}, with $nH = 6.3 \times 10^{21}$ atoms~cm$^{-2}$, $kT = 0.235$ keV, and a normalization of $2.083 \times 10^{-4}$.  The reduced $\chi^2 = 1.33$ for this fit.    
}
\end{figure}

The resulting EPIC spectra of HD 15137 have low signal-to-noise (S/N), but they are consistent with a soft thermal source typical of isolated O-type stars \citep{sana2006}.  We fit the two MOS spectra simultaneously, over the range 0.5--2.3 keV, using a variety of warm absorbed (\textit{wabs}; \citealt{morrison1983}), single temperature (1-T) and two temperature (2-T) thermal models available with Xspec version 11.3.2ag, including Raymond-Smith \citep{raymond1977}, MEKAL \citep{mewe1985, mewe1986, liedahl1995}, APEC \citep{smith2001}, bremsstrahlung \citep{karsas1961, kellogg1975}, and plain blackbody models.  We repeated our fits for the pn spectrum using the same group of models.  A sample 1-T fit of the pn spectrum is shown in Figure \ref{xmmspec}.  The resulting fits were equally good for the 1-T and 2-T models, but a statistical F-test reveals that the second temperature component does not significantly improve the fits.  We also cannot distinguish between the quality of the various 1-T thermal model fits due to the low S/N.  We weakly constrain the temperature to $0.10 \leq kT \leq 0.25$ keV, and the neutral hydrogen column density to $2.8 \times 10^{21} \leq nH \leq 8.6 \times 10^{21}$ atoms~cm$^{-2}$.  There is no evidence of any hard X-ray photons detected in our data.  

Our measured value of $nH$ for HD 15137 can be compared to its reddening, determined from the ultraviolet, optical, and near-infrared spectral energy distribution \citep{mcswain2007a}.  In that work, we measured reddening $E(B-V) = 0.43$ and a ratio of total-to-selective extinction, $R = 3.18$.  The optical extinction is thus $A_V = R \times E(B-V)$.  Several authors have published $nH - A_V$ relations \citep{bohlin1978, wolk2006, guver2009} that predict $nH = 2.5-3.2 \times 10^{21}$ atoms~cm$^{-2}$ based on the measured reddening of HD 15137.  Our measured range in $nH$ is consistent with $A_V$, although our warm absorbed 1-T models also allow values of $nH$ that exceed the expected value by as much as a factor of 3.  

Unfortunately, interpreting our X-ray spectra is made somewhat ambiguous since the emergent spectra of cooling NSs is soft and nearly thermal, very similar to a blackbody \citep{treves2000}.  We performed fits of the EPIC spectra using a model of the hydrogen atmosphere of a NS (\textit{nsa}; \citealt{pavlov1991, zavlin1996}).  As recommended by those authors, we fixed the NS mass to $M_{NS} = 1.4 \; M_\odot$ and the radius to $R_{NS} = 10$ km.  We used a fixed magnetic field strength fixed to $B = 10^{12}$ G and included warm absorption fixed to the predicted $nH = 3 \times 10^{21}$ atoms~cm$^{-2}$ based on the observed $A_V$.  Fitting the two MOS spectra simultaneously results in an unredshifted effective temperature of the NS of $T_{\rm eff, NS} = 437000^{+ 239000}_{- 90000}$ K and a normalization of $1.72 \times 10^{-5}$.  From the fit of the pn spectrum, we obtain $T_{\rm eff, NS} = 955000^{+394000}_{-231000}$ K and a normalization of $7.08 \times 10^{-8}$.  The reduced $\chi^2$ values are equally good as the 1-T models described above, although the NS temperatures from the MOS and pn spectral fits are very inconsistent with each other.   

In order to place an upper limit on any hard power law component that may originate from an accreting compact companion, we repeated the warm absorbed, 1-T models with an additional power law component with photon index $\Gamma = 2$ (fixed).  All of the best fit parameters from the 1-T fits, including nH, $kT$, and their normalizations, were fixed in Xspec.  We then refit each model, allowing only the normalization of the power law component to vary.  In every case, the best fit normalization for the power law component was zero.  We then used the \textit{steppar} routine to investigate the 90\% confidence limit for the power law normalization.  Upon removing the thermal component from the models, we used the fixed $\Gamma$ and the upper limit for its normalization in the remaining absorbed power law model to determine the upper limit for the X-ray flux, $F_X$, of the putative compact object.  Fits of the pn and MOS spectra indicated unabsorbed $F_X \lesssim 10^{-14}$ erg~cm$^{-2}$~s$^{-1}$ (over the range 0.2-10 keV).  At a distance $d = 2.42$ kpc \citep{mcswain2007a}, this corresponds to an X-ray luminosity $L_X \lesssim 10^{31}$ erg~s$^{-1}$ for any hard power law component.


\section{Discussion}

\begin{figure}
\includegraphics[angle=180,scale=0.35]{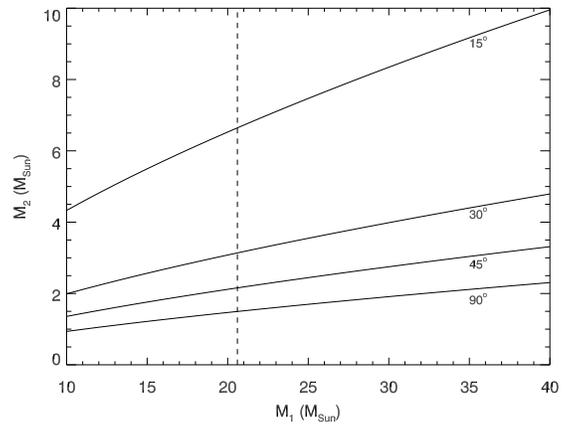}
\caption{
\label{massfcn}
Mass diagram of HD 15137 is plotted for a range of inclination angles (solid lines).  The expected $M_1$ from \citet{martins2005} is also shown (vertical dashed line).  
}
\end{figure}

We can loosely constrain the mass of the companion in HD 15137 using the mass function, $f(m)$, from our new orbital fit.  Assuming the O9.5 III star has a mass of $20.6 \; M_\odot$ \citep{martins2005}, and noting that there is an 87\% probability that the binary inclination $i \ge 30^\circ$, we constrain the companion mass to $1.4 \le M_2 \le 3.0 M_\odot$.  A plot of allowable values of $M_1$ and $M_2$ is shown in Figure \ref{massfcn}.  The most probable range in $M_2$ certainly suggests that a low mass NS or BH could present in the system.  The kick velocity from a prior supernova event is a reasonable explanation for the current high peculiar space velocity ($63 \pm 12$ km~s$^{-1}$; \citealt{mcswain2007a}) for the system.  A white dwarf companion is unlikely due to the initial mass of the supernova progenitor.

HD 15137 has a measured mass loss rate, $\log \dot{M} = -6.5$ $M_\odot$~yr$^{-1}$ \citep{howarth1989} and a terminal wind velocity, $v_\infty = 1690$ km~s$^{-1}$ \citep{mcswain2007b}.  Using these wind parameters and our new orbital solution, we can predict the X-ray luminosity emitted from a possible wind-accreting NS in the system using a simple Bondi-Hoyle wind accretion model \citep{lamers1976}.  With the O star mass and a NS companion with mass $1.4 \; M_\odot$, we should expect to observe an unabsorbed $F_X \sim 2 \times 10^{-12}$ erg~cm$^2$~s$^{-1}$ (corresponding to an unabsorbed X-ray luminosity $\log L_X = 33.2$ erg~s$^{-1}$) at $\phi = 0.7$ if wind accretion was indeed present.  A more massive, accreting BH should result in a proportionately higher $L_X$.  However, the \textit{XMM-Newton} observations rule out such a bright X-ray source by at least two orders of magnitude, clearly ruling out an accreting BH companion.  O-type stars typically exhibit order-of-magnitude variations in their mass-loss rates \citep{mcswain2004}, so if the mass loss rate has decreased with time since measured by \citet{howarth1989}, the expected wind accretion luminosity should decrease proportionately.  However, variability in $\dot{M}$ is not sufficient to explain the discrepancy between the predicted $F_X$ and our observed upper limit.  

The dynamical age of the putative NS in HD 15137 also allows us to discriminate against the cooling NS model.  A young NS cools to a temperature $\sim 10^6$ K after about $10^3$--$10^4$ yr after its birth, and it is expected to remain at that roughly constant temperature for $\sim 10^5$ yr \citep{treves2000}.  Since the binary was ejected $10^7$ yr ago from the open cluster of its birth \citep{boyajian2005}, then it should have cooled to temperatures far below our measured $T_{\rm eff, NS}$.  

If there is a NS or BH companion present in HD 15137, it is more likely in a quiescent state.  Quiescent NSs in the propellor regime have been observed with hard X-ray spectra with $\Gamma \sim 2$ and unabsorbed luminosities of $33.5 \lesssim \log L_X \lesssim 37.5$ erg~s$^{-1}$ \citep{campana2002}.  Our upper limits on the hard X-ray flux of HD 15137 clearly rule out such a NS in the propellor regime.  NSs in the ejector regime are potentially observed as radio pulsars, but our pulsar search did not detect any such candidate \citep{mcswain2007b}.  We cannot firmly rule out a NS companion in HD 15137, but our observations suggest that a companion is unlikely to be present.  However, a quiescent BH is quite possible.  Black hole X-ray binaries spend most of their lives in quiescence with an X-ray luminosity as little as $10^{-6}$ of an active accretor \citep{pszota2008, coriat2009}, consistent with our upper limits on $L_X$.  

If HD 15137 has a ``twin'' companion with mass ratio $q = M_2/M_1 \sim 0.95$ \citep{pinsonneault2006, lucy2006}, the system must have an improbably low $i \sim 6^\circ$.  The late O-type or early B-type companion would have spectral lines heavily blended with the primary star, making it difficult to detect from $V_r$ variations but possibly identifiable using doppler tomography.  
\citet{boyajian2005} failed to detect an optical companion in HD 15137 using doppler tomography, but we repeated such an effort here with the advantage of a new orbital solution for the binary.  We assumed a wide range of $0.05 \le q \le 0.95$ in an effort to detect a massive or low mass secondary.  The reconstructed spectra from doppler tomography do a poor job of matching realistic H Balmer line profiles, and all possible reconstructed spectra maintain constant \ion{He}{1}:\ion{He}{2} line strength ratios.  These line ratios would be expected to differ from the primary star in any cooler optical companion, so we cannot claim any detection of a possible optical secondary star with the tomography.  


\citet{boyajian2005} noted that the travel time since the ejection of HD 15137 presents an unusual paradox since it is longer than the expected lifetime of an O-type star.  \citet{perets2009} offers a possible explanation:  the close binary may have been ejected from its parent cluster by a dynamical ejection, while later mass transfer rejuvenated the massive star and extended its lifetime.  The orbital period and probable high mass ratio of HD 15137 are consistent with the strong mass transfer scenario \citep{perets2009}.  However, with a total binary mass almost certainly $> 20 M_\odot$, the binary might have originally been composed of two $\sim 10 M_\odot$ stars.  Perets predicts a very small parameter space for such massive binaries that will experience strong mass transfer; most of the massive binaries in their simulations experience weak mass transfer or mergers.  While rejuvenation through mass transfer is possible, it seems an unlikely explanation for the current properties of HD 15137.  On the other hand, it is also possible that HD 15137 experienced a supernova after the binary's ejection via dynamical processes.  In such case, the age of the compact companion would not correspond to the dynamical age found by \citet{boyajian2005}, and a young, cooling NS companion might be present.  However, such a two-step evolutionary scenario would obscure the true dynamical history of HD 15137; any supernova kick velocity would alter the system's trajectory and render any measurement of its travel time unreliable.  It is impossible to draw any conclusions regarding the dynamical ejection and subsequent evolutionary history of HD 15137.


\section{Summary and Conclusions}

We have searched for a compact companion in the massive runaway binary HD 15137, to no avail.  \textit{XMM-Newton} EPIC spectra of the system are consistent with a soft thermal source typical of isolated O-type stars \citep{sana2006}.  The dynamical age of HD 15137 implies the system is too old to contain a cooling NS with $T_{\rm eff, NS} \sim 10^5 -10^6$ K, although our X-ray spectra are consistent with such a NS.  Our lack of detection of hard X-ray photons is inconsistent with the expected emission from a quiescent NS in the propellor regime or a stellar wind-accreting NS, but a quiescent BH remains a possibility.  

Similarly, we have been unable to identify an optical secondary in the system.  Doppler tomography based on our updated orbital solution was unsuccessful, and spectral modeling of the primary is consistent with a single O9.5 III star.  Weak N enrichment and C depletion are observed in the optical spectra, possible evidence of CNO processed gas at the stellar surface.  The companion probably has a mass in the range 1--3 $M_\odot$, implying a very low mass ratio of $q < 0.2$.  For a MS companion, this would correspond to a late B star which would contribute a negligible flux compared to the bright O-type star.  Such an optical companion would be difficult if not impossible to detect in the close, eccentric orbit of HD 15137.  

Although it might be tempting to rule out a compact companion based on the lack of its detection in our \textit{XMM-Newton} observation, it is difficult to imagine how the binary might have been ejected as a runaway with a low mass stellar companion in a close gravitational encounter.  HD 15137 bears many similarities with the class of Be/X-ray biniaries (BeXRB): a moderately long orbital period, high eccentricity, low mass function, and a rapidly rotating primary star.  Although \citet{boyajian2005} reported no H$\alpha$ emission from HD 15137, the  rapid rotation and possible nonradial pulsations make the star a good candidate for a transient Be star that experiences sporadic disk disappearances \citep{mcswain2009}.  Many BeXRB are X-ray transients that are invisible in X-rays except at times when the NS encounters higher density disk gas (there are no known BH companions in BeXRB systems; \citealt{belczynski2009}).  X-ray outbursts may happen either when a Be star disk outburst occurs and the disk density increases close to the NS, or when the NS crosses the plane of the disk (in cases where the orbital and disk planes are different).  The \textit{XMM-Newton} observations may have occurred when conditions were unfavorable for accretion.


\acknowledgments

We thank the referee, Hagai Perets, for suggestions that improved this manuscript.  We are grateful to the staff at KPNO for their hard work to schedule and support these observations.  
MD would like to thank Drs.\ G.\ Rauw and N.\ Linder for taking some of the OHP spectra, the staff of OHP for the technical support during the various observing runs, and the Minist\`ere de l'Enseignement Sup\'erieur et de la Recherche de la Communaut\'e Fran\c{c}aise de Belgique for financial support for the OHP observing runs.  MVM and CA would also like to thank the Wyoming Infrared Observatory for providing time to finish this manuscript and for a renewed appreciation of the small things in life.  This work is supported by NASA DPR numbers NNX08AX79G and NNG08E1671 and an institutional grant from Lehigh University.

{\it Facilities:} \facility{KPNO:CFT}, \facility{OHP:Aur\'elie}, \facility{XMM:EPIC}.



\clearpage
\begin{deluxetable}{lccc}
\tablewidth{0pt}
\tablecaption{Radial Velocity Measurements of HD 15137\label{vr}}
\tablehead{
\colhead{HJD} &
\colhead{ } &
\colhead{$V_r$} &
\colhead{$\sigma$} \\
\colhead{($-$2,400,000)} &
\colhead{Telescope} &
\colhead{(km s$^{-1}$)} &
\colhead{(km s$^{-1}$)} }
\startdata
53652.642 \dotfill  &  OHP  &  $-39.4$  &  \phn  3.8  \\
53654.538 \dotfill  &  OHP  &  $-37.5$  &  \phn  2.4  \\
53980.585 \dotfill  &  OHP  &  $-55.7$  &  \phn  2.6  \\
53982.552 \dotfill  &  OHP  &  $-42.7$  &  \phn  2.2  \\
53984.545 \dotfill  &  OHP  &  $-43.5$  &  \phn  1.8  \\
54033.406 \dotfill  &  OHP  &  $-53.6$  &  \phn  9.3  \\
54033.495 \dotfill  &  OHP  &  $-56.9$  &       10.3  \\
54033.602 \dotfill  &  OHP  &  $-54.4$  &  \phn  8.8  \\
54034.670 \dotfill  &  OHP  &  $-56.5$  &  \phn  9.2  \\
54035.413 \dotfill  &  OHP  &  $-58.0$  &  \phn  1.3  \\
54035.437 \dotfill  &  OHP  &  $-59.8$  &  \phn  4.9  \\
54035.458 \dotfill  &  OHP  &  $-56.9$  &  \phn  8.8  \\
54035.480 \dotfill  &  OHP  &  $-59.8$  &  \phn  6.1  \\
54035.502 \dotfill  &  OHP  &  $-54.4$  &       10.0  \\
54035.524 \dotfill  &  OHP  &  $-57.2$  &  \phn  6.7  \\
54035.545 \dotfill  &  OHP  &  $-59.6$  &  \phn  2.8  \\
54035.568 \dotfill  &  OHP  &  $-59.6$  &  \phn  3.2  \\
54035.588 \dotfill  &  OHP  &  $-56.2$  &  \phn  5.2  \\
54035.611 \dotfill  &  OHP  &  $-53.5$  &  \phn  4.4  \\
54035.632 \dotfill  &  OHP  &  $-56.2$  &  \phn  5.2  \\
54035.655 \dotfill  &  OHP  &  $-52.4$  &  \phn  6.3  \\
54396.439 \dotfill  &  OHP  &  $-29.1$  &  \phn  5.9  \\
54397.516 \dotfill  &  OHP  &  $-31.5$  &  \phn  3.5  \\
54400.646 \dotfill  &  OHP  &  $-31.7$  &  \phn  4.6  \\
54401.598 \dotfill  &  OHP  &  $-31.6$  &  \phn  5.9  \\
54402.588 \dotfill  &  OHP  &  $-28.4$  &  \phn  5.2  \\
54405.599 \dotfill  &  OHP  &  $-25.4$  &  \phn  5.2  \\
54406.590 \dotfill  &  OHP  &  $-22.8$  &  \phn  2.6  \\
54407.532 \dotfill  &  OHP  &  $-33.6$  &  \phn  1.7  \\
54408.591 \dotfill  &  OHP  &  $-29.1$  &  \phn  4.0  \\
54409.527 \dotfill  &  OHP  &  $-29.7$  &  \phn  4.2  \\
54410.475 \dotfill  &  OHP  &  $-23.5$  &  \phn  2.7  \\
54411.551 \dotfill  &  OHP  &  $-27.2$  &  \phn  8.0  \\
54412.504 \dotfill  &  OHP  &  $-28.6$  &  \phn  7.9  \\
54413.584 \dotfill  &  OHP  &  $-35.9$  &  \phn  6.0  \\
54414.530 \dotfill  &  OHP  &  $-31.1$  &  \phn  7.0  \\
54415.545 \dotfill  &  OHP  &  $-34.8$  &  \phn  7.5  \\
54416.453 \dotfill  &  OHP  &  $-30.3$  &  \phn  4.6  \\
54417.467 \dotfill  &  OHP  &  $-32.6$  &  \phn  6.1  \\
54418.383 \dotfill  &  OHP  &  $-33.3$  &  \phn  6.8  \\
54419.412 \dotfill  &  OHP  &  $-45.3$  &  \phn  5.4  \\
54421.525 \dotfill  &  OHP  &  $-54.5$  &  \phn  1.1  \\
54422.420 \dotfill  &  OHP  &  $-56.4$  &  \phn  1.8  \\
54423.322 \dotfill  &  OHP  &  $-51.0$  &  \phn  3.6  \\
54756.820 \dotfill  &  CF   &  $-73.8$  &       17.9  \\
54757.753 \dotfill  &  CF   &  $-57.3$  &  \phn  7.7  \\
54758.831 \dotfill  &  CF   &  $-52.0$  &  \phn  4.1  \\
54759.771 \dotfill  &  CF   &  $-51.1$  &  \phn  4.2  \\
54760.781 \dotfill  &  CF   &  $-50.5$  &  \phn  2.1  \\
54761.815 \dotfill  &  CF   &  $-50.6$  &  \phn  3.2  \\
54762.773 \dotfill  &  CF   &  $-39.8$  &  \phn  4.7  \\
54763.779 \dotfill  &  CF   &  $-48.0$  &  \phn  5.8  \\
54764.761 \dotfill  &  CF   &  $-48.0$  &  \phn  2.1  \\
54765.741 \dotfill  &  CF   &  $-45.2$  &  \phn  1.1  \\
54766.730 \dotfill  &  CF   &  $-46.0$  &  \phn  4.5  \\
54767.955 \dotfill  &  CF   &  $-47.3$  &  \phn  5.3  \\
54768.697 \dotfill  &  CF   &  $-49.0$  &       11.2  \\
54769.721 \dotfill  &  CF   &  $-48.4$  &  \phn  4.0  \\
54771.663 \dotfill  &  CF   &  $-39.4$  &  \phn  1.4  \\
54771.674 \dotfill  &  CF   &  $-41.1$  &  \phn  6.5  \\
54771.685 \dotfill  &  CF   &  $-36.7$  &  \phn  1.4  \\
54771.696 \dotfill  &  CF   &  $-37.4$  &  \phn  3.5  \\
54771.712 \dotfill  &  CF   &  $-40.3$  &  \phn  5.2  \\
54771.722 \dotfill  &  CF   &  $-36.4$  &  \phn  4.8  \\
54771.733 \dotfill  &  CF   &  $-39.5$  &  \phn  5.3  \\
54771.744 \dotfill  &  CF   &  $-39.6$  &  \phn  2.7  \\
54771.754 \dotfill  &  CF   &  $-38.1$  &  \phn  2.9  \\
54771.765 \dotfill  &  CF   &  $-34.4$  &  \phn  5.3  \\
54771.776 \dotfill  &  CF   &  $-38.9$  &  \phn  5.0  \\
54771.786 \dotfill  &  CF   &  $-40.8$  &  \phn  2.7  \\
54771.797 \dotfill  &  CF   &  $-43.3$  &  \phn  2.9  \\
54771.808 \dotfill  &  CF   &  $-37.3$  &  \phn  6.7  \\
54771.818 \dotfill  &  CF   &  $-43.3$  &  \phn  4.9  \\
54771.829 \dotfill  &  CF   &  $-45.8$  &  \phn  4.2  \\
54772.641 \dotfill  &  CF   &  $-40.7$  &  \phn  7.5  \\
54774.751 \dotfill  &  CF   &  $-35.9$  &  \phn  4.8  \\
54775.702 \dotfill  &  CF   &  $-45.9$  &  \phn  3.8  \\
54777.696 \dotfill  &  CF   &  $-38.4$  &       11.4  \\
54779.754 \dotfill  &  CF   &  $-43.9$  &  \phn  2.4  \\
54780.886 \dotfill  &  CF   &  $-32.8$  &       10.5  \\
54781.654 \dotfill  &  CF   &  $-36.5$  &       11.6  \\
54782.629 \dotfill  &  CF   &  $-35.4$  &  \phn  4.2  \\
54783.659 \dotfill  &  CF   &  $-43.1$  &  \phn  5.0  \\
54784.654 \dotfill  &  CF   &  $-39.3$  &  \phn  2.9  \\
54785.630 \dotfill  &  CF   &  $-43.0$  &  \phn  5.3  \\
54786.600 \dotfill  &  CF   &  $-47.8$  &  \phn  4.5  \\
54787.660 \dotfill  &  CF   &  $-44.3$  &  \phn  6.2  \\
54788.695 \dotfill  &  CF   &  $-42.0$  &  \phn  6.1  \\
54789.907 \dotfill  &  CF   &  $-35.8$  &  \phn  2.2  \\
54790.909 \dotfill  &  CF   &  $-34.0$  &       10.8  \\
54791.943 \dotfill  &  CF   &  $-32.0$  &       10.3  \\
\enddata
\end{deluxetable}

\begin{deluxetable}{lc}
\tablewidth{0pt}
\tablecaption{Orbital elements of HD 15137 \label{orbit}}
\tablehead{
\colhead{Element} &
\colhead{Value} }
\startdata
$P$ (d)                      		& \phs\phn\phn $55.3957 \pm 0.0038$ \phn				\\
$T_0$ (HJD -- 2,400,000) 	& \phs\phn $54421.991 \pm 0.064$ \phn\phn\phn				\\
$K$ (km s$^{-1}$) 		& \phs\phn\phn\phn\phn $13.56 \pm 0.15$  \phn\phn\phn		\\
$V_0$ (km s$^{-1}$)		& \phn\phn\phn\phn $-42.06 \pm 0.11$ \phn\phn\phn		\\
$e$					& \phs\phn\phn\phn\phn $0.6239 \pm 0.0088$ \phn\phn		\\
$\omega$ ($^\circ$)		& \phs\phn\phn\phn $152.17 \pm  0.86$ \phn\phn\phn	 	\\
$f(M_2)$ ($M_\odot$) 	& \phs\phn\phn\phn $0.00685 \pm 0.00030$ \phn				\\
$a_1 \sin i$ ($R_\odot$)	& \phs\phn\phn\phn $11.60 \pm 0.17$  \phn\phn			\\
r.m.s. (km s$^{-1}$)		& 4.28                   \\
\enddata
\end{deluxetable}
\begin{deluxetable}{lcccc}
\tablewidth{0pt}
\tablecaption{He Line Ratios and Effective Temperature \label{specmodel}}
\tablehead{
\colhead{ } &
\colhead{ } &
\colhead{$T_{\rm eff}$ (K)} &
\colhead{$T_{\rm eff}$ (K)} &
\colhead{$T_{\rm eff}$ (K)} \\
\colhead{ } &
\colhead{ $\log \frac{W_\lambda({\rm He II})}{W_\lambda({\rm He I})}$ } &
\colhead{($\log g$=3.25)} &
\colhead{($\log g$=3.50)} &
\colhead{($\log g$=3.75)} }
\startdata
Using OSTAR2002 grid:  &  &  &  & \\
\ion{He}{2} $\lambda4542$:\ion{He}{1} $\lambda4471$ \dotfill & $-0.656$ & 27041 & 27814 & 28719 \\
\ion{He}{2} $\lambda4542$:\ion{He}{1} $\lambda4388$ \dotfill & $-0.295$ & 28695 & 30153 & 30971 \\
\ion{He}{2} $\lambda4200$:\ion{He}{1} $\lambda4144$ \dotfill & $-0.122$ & 30099 & 31202 & 32390 \\
Mean $T_{\rm eff}$ \dotfill & \nodata &  28600 & 29700 & 30700 \\
Reduced $\chi^2$ from CF spectrum \dotfill & \nodata & 11.0 & 2.1 & 9.6 \\
Reduced $\chi^2$ from OHP spectrum \dotfill & \nodata & 4.3 & 1.4 & 7.9 \\
\\
Using BSTAR2006 grid:  &  &  &  &  \\
\ion{He}{2} $\lambda4542$:\ion{He}{1} $\lambda4471$ \dotfill & $-0.656$  & 27764 & 28889 & 29954 \\
\ion{He}{2} $\lambda4542$:\ion{He}{1} $\lambda4388$ \dotfill & $-0.295$  & 29983 & 31527 & 33532 \\
\ion{He}{2} $\lambda4200$:\ion{He}{1} $\lambda4144$ \dotfill & $-0.122$  & 31640 & 33798 & 36645 \\
Mean $T_{\rm eff}$ \dotfill & \nodata &  29800 & 31400 & 33400 \\
Reduced $\chi^2$ from CF spectrum \dotfill & \nodata & 2.8 & \nodata & \nodata \\
Reduced $\chi^2$ from OHP spectrum \dotfill & \nodata & 1.7 & \nodata & \nodata \\
\enddata
\end{deluxetable}

\begin{deluxetable}{lccccc}
\tabletypesize{\scriptsize}
\tablecaption{Journal of \textit{XMM-Newton} observations of HD 15137  \label{xmmjournal}}
\tablewidth{0pt}
\tablehead{
\colhead{ } &
\colhead{Time of Mid Exposure} &
\colhead{Performed} &
\colhead{Effective} &
\colhead{Source Count} &
\colhead{Background Count} \\
\colhead{Camera} &
\colhead{(JD$-$2,400,000)} &
\colhead{Duration (ks)} &
\colhead{Duration (ks)} &
\colhead{Rate (cnt s$^{-1}$)} &
\colhead{Rate (cnt s$^{-1}$)} }
\startdata
MOS1	&  54681.56704  &  19.667  &  15.90  &  $0.0100 \pm 0.0008$  &  $0.0036 \pm 0.0008$ 	\\
MOS2	&  54681.56741  &  19.672  &  17.50  &  $0.0069 \pm 0.0006$  &  $0.0014 \pm 0.0007$	\\
pn	&  54681.57333  &  18.032  &  18.10  &  $0.134 \pm 0.003$    &  $0.018 \pm 0.003$ 	\\
\enddata
\end{deluxetable}



\clearpage

\clearpage

\end{document}